\documentclass[aps,pra,groupedaddress,10pt]{revtex4-2}
\usepackage{amsmath}
\usepackage{amssymb}

\begin{document}


\title{On the Non-Flatness Nature of Noncommutative Minkowski Spacetime and the Singular Behavior of Probes}


\author{Manali Roy}
\email[]{manaliroy37@pondiuni.ac.in}
\author{Muthukumar Balasundaram}
\email[]{muthukbs@pondiuni.ac.in}
\affiliation{Department of Physics, School of Physical Chemical and Applied Sciences, Pondicherry University, Pondicherry-605014, India.}


\date{\today}

\begin{abstract}
It is more than a century-old concept that the Minkowski spacetime is flat. From the pure geometric point of view, we explicitly address the issue of whether a noncommutative Minkowski spacetime is flat or not. In the framework of the twisted-diffeomorphism approach to noncommutative gravity with canonical type noncommutative (NC) coordinate structure, one important result that we get is that the NC Minkowski spacetime parametrized either with spherical polar coordinates or with parabolic coordinates has nontrivial NC corrections to Riemann curvature tensor, Ricci tensor and curvature scalars. Another crucial result is that the curvature scalars have singular behavior at certain points, and these singularities are not coordinate singularities.  The nature of these singularities clearly points towards the idea of high-energetic probes turning into black holes. The absence of any such noncommutative corrections, and thus any such singularities, in the cases of Cartesian coordinates and cylindrical coordinates leads to the conclusion that high-energetic probes do not turn into black holes when the canonical NC structure is considered for these coordinates. 
\end{abstract}

\maketitle

%
\section{\label{sec:level1}Introduction:}
In the Planck length scale, the concept of spacetime as continuous manifold breaks down, and it is expected to have the mathematical structure of quantum spacetime\cite{Szabo:2001kg}. As the spacetime is quantized, the spacetime coordinates are replaced by Hermitian operators, which do not commute among themselves. So a noncommutative spacetime is expected to play a significant role in the physics of the Planck length scale \cite{Meljanac:2017jyk,Besnard:2020lto,Piscicchia:2022xra,Frob:2022ciq,Smith:2022bwv,Piscicchia:2022eod,Arzano:2022vmh}. In addition to the possibility of a minimal length associated with such noncommutative (NC) spacetimes, the quantization of length also shows up in certain cases \cite{Balasundaram:2022esk}. On the one hand, the idea of spacetime noncommutativity emerges from the attempts to unify the gravity and the quantum \cite{Doplicher:1994zv,Doplicher:1994tu}, the emergent gravity phenomena in noncommutative gauge field theories, on the other hand, take the quantum field theories a step closer to bring the gravitational effects into the quantum field theories \cite{Gross:2000ph}-\cite{Nandi:2021xtr}, and in the same vein, the gravitational theories in NC spacetimes are expected to bring the quantum effects into the gravitational theories.

There are a number of different approaches to NC gravity. With the idea of the Seiberg-Witten map \cite{Seiberg:1999vs} being extended to enveloping algebra \cite{Jurco:2000ja}, gauge theories of gravity in NC spacetime have been studied by mapping the commutative gauge theories of gravitation to their NC counterparts. In one approach, the noncommutative SO(4,1) de sitter group is gauged and contracted to the Poincare group ISO(3,1) using the Seiberg-Witten map \cite{Chaichian:2007we}. Using the concept of twisted Poincare algebra\cite{Chaichian:2004za,Chaichian:2004yh}  and extending the idea to general coordinate transformations, gravity theory in NC spacetime is also constructed \cite{Aschieri:2005yw}. When commutation relations of the spacetime coordinates have a Lie Algebra structure, there is a class of volume-preserving general coordinate transformation which preserves the symmetry of the NC algebra \cite{Chaichian:2007we,Calmet:2005qm,Calmet:2006iz,Harikumar:2006xf}. The theory of gravity corresponding to volume-preserving diffeomorphism is a unimodular theory of gravity. In this case, NC  general theory of relativity is constructed by gauging the Lorenz gauge group SO(3,1) in the enveloping algebra framework. Two-dimensional NC gravity theories have also been proposed based on NC gauge theories \cite{Balachandran:2006qg}. 

Since the general theory of relativity is invariant under diffeomorphism, a theory of NC gravity is formulated using the mathematical framework of twist-deformed diffeomorphism \cite{Aschieri:2005yw}. In this theory of gravitation, which is invariant under deformed diffeomorphism, the Christoffel symbols, Riemann curvature tensor, Ricci tensor and curvature scalar, etc., are mathematically constructed through star products in NC spacetime, and the effect of noncommutation of spacetime coordinates manifests as the NC  corrections to these geometrical elements.

The idea behind the proposal of NC spacetime is that if a highly energetic probe is used to explore the Planck length scale, the probe will turn into a black hole, forming a tiny volume within its event horizon beyond the reach of any experimental observation \cite{Doplicher:1994zv, Doplicher:1994tu,Szabo:2009tn}, and thereby creating a volume of uncertainty. This volume of uncertainty is mimicked using an NC structure among rectilinear spacetime coordinates, which introduces a short-length scale to avoid singularity. 

Within the framework of noncommutative formalism, it has not been shown in the literature which types of NC structures of coordinates lead to singularities associated with probes and which do not lead to. The aim of this article is to do this analysis which is required because an NC structure among cylindrical coordinates turns out to be a viable alternative if the idea is to avoid any singularity associated with a probe. Our work considers the twist-deformed diffeomorphism framework of NC gravity in canonical NC spacetime, and our analysis also shows that such singularities do arise if we twist the theories on flat Minkowski spacetime with the Drinfel'd twist  $e^{\frac{i\Theta^{\mu\nu}}{2}\frac{\partial}{\partial x^\mu} \otimes \frac{\partial}{\partial x^\nu}}$, where $\{x^\mu\}$ is either the set of spherical polar coordinates $(r,\theta,\phi)$ and time or the set of the parabolic coordinates $(u,v,\phi)$ and time, with $\Theta^{\mu\nu}$ being a constant in either of the cases. But if $\{x^\mu\}$ is the set of Cartesian/cylindrical coordinates and time, then such singularities associated with probes do not arise. 

The work is organized as follows. We briefly review the idea of twist-deformed diffeomorphism in Section \ref{sec:ncintro}. The key point in Section \ref{sec:cartcyl} is that when the metric tensor components are functions of at most one coordinate, there are no NC corrections to curvature tensors and curvature scalars. The NC effects occur in the pure gravity/geometry sector only if the metric tensor depends on at least two coordinates that do not commute. The reason is discussed in Section \ref{sec:tetrads}. Section \ref{sec:scalarinv} contains some remarks on the NC scalar invariants. In Section \ref{sec:ncmspheri}, we deal with the NC curvature corrections to NC Minkowski spacetime parametrized by spherical polar coordinates. We have done both first- and second-order noncommutative corrections to inverse metric tensor, Christoffel symbols, Riemann curvature tensor, Ricci tensor, Ricci scalar and Kretschmann scalar in the twist-deformed diffeomorphism framework of NC gravity. We repeat these calculations in Section \ref{sec:ncmpara} for the NC Minkowski spacetime parametrized by parabolic coordinates since this is another example of a metric tensor with more than one coordinate dependency and with nontrivial NC curvature corrections. The concluding remarks are given in Section \ref{sec:conclu}. 

One main result of our calculations is that the Riemann curvature tensor, Ricci tensor and the curvature scalar acquire nontrivial corrections in the case of NC Minkowski spacetime parametrized either by spherical polar or by parabolic coordinates. 
This turns out to be a general feature of spacetime in which the metric tensor is a function of at least two coordinates that are promoted as noncommuting operators. Another main result is that the nontrivial corrections to curvature tensors and curvature scalars have singularities at certain points. The nature of these singularities is such that they should be associated with length-measuring probes rather than with spacetime as such. Also, they are not coordinate singularities and they can be avoided in these coordinate systems by an appropriate choice of the NC parameter $\Theta^{\mu\nu}$. 
\section{\label{sec:ncintro}Deformed Diffeomorphism on NC Spacetime}
There is a one-to-one correspondence between a Hausdorff topological space and the commutative algebra of complex-valued functions defined on it with a pointwise multiplication rule governing the functions. But an NC spacetime is related to an NC algebra of functions defined on it. An NC spacetime is characterized by an NC algebra of twisted-diffeomorphism where the pointwise multiplication rule is replaced by a deformed star-product of the complex-valued functions defined on it. An NC gravity theory on NC spacetime is invariant under twisted-diffeomorphism \cite{Aschieri:2005yw} in the same way that a gravity theory in a Riemannian spacetime is invariant under diffeomorphism, 

The following commutation relations among spacetime coordinates define an NC spacetime:
\begin{align}
 \left[\hat{x}^{\mu},~\hat{x}^{\nu}\right]=i\Theta^{\mu\nu},
\end{align}
where $\Theta_{\mu\nu}$ is a constant and real antisymmetric matrix. The above operator algebra can be realized on a linear space of complex functions $f(x)$ of commuting variables if the algebra of functions is given a noncommutative multiplication structure called Moyal-Weyl star-product (*-product) \cite{Bayen:1977ha}: 
\begin{align}
 f(x) *g(x)=\displaystyle \left. e^{ \frac{i}{2}\Theta^{\mu\nu}\frac{\partial}{\partial x^{\mu}}\frac{\partial}{\partial x'^{\nu}}} f(x)g(x')\right|_{x'=x}.
\end{align}
The concept of deformed diffeomorphism relies on the idea that the general coordinate transformations in general relativity are, in general, diffeomorphism transformations. If $\delta_{\xi}$ denotes the infinitesimal change in a scalar field $\phi(x)$ under the infinitesimal general coordinate transformation $x^{\mu}\to x'^{\mu}=x^{\mu}+\xi^{\mu}(x)$, then the invariance relation $\phi'(x')=\phi(x)$ implies the field transformation $\phi(x) \to \phi'(x)=\phi(x) + \delta_{\xi} \phi=\phi(x)-\xi^{\mu}(x)\partial_{\mu}\phi$. Also, for the pointwise multiplication of two transformed fields, the following equality holds:
\begin{align}
\phi'(x)\cdot \chi'(x)=\phi(x)\cdot \chi(x)+\delta_{\xi}\left(\phi(x)\cdot \chi(x)\right)
\end{align}
because of the Leibniz rule
\begin{subequations}\label{eq-leibniz}
\begin{align}
(\delta_{\xi}\phi )\cdot \chi + \phi \cdot(\delta_{\xi} \chi) = 
\delta_{\xi}(\phi\cdot \chi). \label{eq-leibniza}
\end{align}
If $\mu$ denotes the map that maps a tensor product to the pointwise 
multiplication of commutative algebra, i.e., $\mu(\phi\otimes\chi)=\phi\cdot 
\chi$, then using the coproduct $\triangle(\delta_{\xi}) = \delta_{\xi}\otimes 1 
+ 1  \otimes \delta_{\xi}$,  Eq.\eqref{eq-leibniza} from right to left can 
be written as 
\begin{align}
\begin{aligned} \label{eq-leibnizb}
\delta_{\xi} \mu\{\phi\otimes\chi\} =
\mu\{\triangle(\delta_{\xi})(\phi\otimes\chi)\}.
\end{aligned}
\end{align}
\end{subequations}
The key point in the above Eq.(\ref{eq-leibnizb}) is the commutation relation 
$\delta_{\xi}\mu=\mu\triangle(\delta_{\xi})$. For three successive infinitesimal 
transformations involving $-\xi, \eta$ and $\xi$, the commutator 
$[\delta_{\xi},\delta{\eta}]=\delta_{\xi\times\eta}$ implies the closure 
property of the algebra of vector fields $\xi^{\mu}\partial_{\mu}$ acting on a 
differential manifold. Essentially, the commutator  denotes the algebra of 
diffeomorphism, and its coproduct-counterpart is worked out to be 
$[\triangle(\delta_{\xi}),\triangle(\delta_{\eta})]=\triangle(\delta_{
\xi\times\eta})$. The Leibniz rule \eqref{eq-leibniz} holds regardless of 
whether $\delta_{\xi}$, represented by $-\xi^{\mu}(x)\partial_{\mu}$, is independent of $x$ or not. But if $\delta_{\xi}$ depends on $x$ and the pointwise product of 
the fields is replaced by *-product, then Eq.\eqref{eq-leibniz} does not 
hold, i.e., $ (\delta_{\xi}\phi )* \chi + \phi* (\delta_{\xi}\chi) \neq 
\delta_{\xi}(\phi*\chi)$. This is exactly where the concept of twist-deformed 
diffeomorphism comes into the picture. 
If $\mu_*$ denotes the mapping of tensor product to the *-product of NC coordinate algebra, i.e., 
\begin{align}
\mu_*\{\phi\otimes\chi\}=\mu \mathcal{F}\{\phi\otimes\chi\}=\phi*\chi, 
\end{align}
where $\mathcal{F}=e^{ \frac{i}{2}\Theta^{\mu\nu}\frac{\partial}{\partial x^{\mu}}\otimes\frac{\partial}{\partial x^{\nu}}}$ is called the Drinfel'd twist, then using the twisted coproduct defined by 
\begin{align}
\begin{aligned}
\triangle_t(\delta_{\xi})&= e^{ -\frac{i}{2}\Theta^{\mu\nu}\frac{\partial}{\partial x^{\mu}}\otimes\frac{\partial}{\partial x^{\nu}}}\left(\delta_{\xi}\otimes 1 + 1  \otimes \delta_{\xi}\right)e^{ \frac{i}{2}\Theta^{\mu\nu}\frac{\partial}{\partial x^{\mu}}\otimes\frac{\partial}{\partial x^{\nu}}}\\
&=\mathcal{F}^{-1}\left(\delta_{\xi}\otimes 1 + 1  \otimes \delta_{\xi}\right) \mathcal{F},
\end{aligned}
\end{align}
the analogue of \eqref{eq-leibnizb} in noncommutative algebra can be written 
as 
\begin{align}
\begin{aligned}
 \delta_{\xi}u_*\{\phi\otimes\chi \} & =
 &\mu_*\{\triangle_t(\delta_{\xi})(\phi\otimes\chi)\}, 
 \label{eq-leibnizt}
 \end{aligned}
\end{align}
where an $\mathcal{F}\mathcal{F}^{-1}(=1)$ has been inserted on the right side. In order to take the $\delta_{\xi}$ out of the tensor product on the right side of Eq.\eqref{eq-leibnizt}, to the left of $\mu_*$, the coproduct needs to be twisted: 
$\mu_*\triangle_t(\delta_{\xi})=\delta_{\xi}\mu_*$. The above rule 
Eq.\eqref{eq-leibnizt} is easily generalized to the product of arbitrary 
tensor fields leading to a covariant construction of theories.

Essentially the Drinfel'd twist $\mathcal {F}=e^{ \frac{i}{2}\Theta^{\mu\nu}\frac{\partial}{\partial x^{\mu}}\otimes\frac{\partial}{\partial x^{\nu}}}$ introduces noncommutation among the coordinates. Drinfel'd twist satisfies the cocycle condition $(\mathcal{F}\otimes 1) (\triangle\otimes id) \mathcal{F}=(1\otimes \mathcal{F}) (id \otimes\triangle)\mathcal{F}$ responsible for the associative property of the deformed product. It also satisfies the normalization condition $(\epsilon\otimes id)\mathcal{F}=1\otimes 1 =(id \otimes\epsilon)\mathcal{F}$. In this twisted-diffeomorphism approach to NC gravity, the basic element is the classical vierbein $e^{~a}_{\mu}(x)$, also called tetrad in 3+1 dimensions, and this is taken to be the full NC version of vierbein to all orders in the NC  parameter $\Theta$ \cite{Aschieri:2005yw}. But the metric in NC spacetime is defined to have the following symmetric form with $\Theta$-dependency:
\begin{equation} 
G_{\mu\nu}=\frac{1}{2}\left(e^{~a}_{\mu}*e^{~b}_{\nu}+e^{~a}_{\nu}*e^{~b}_{\mu
} \right)\,\eta_{ab}, \label{eq-ncmetric}
\end{equation}
with its zeroth order being the metric in the commutative case, i.e., $G^{(0)}_{\mu\nu}=g_{\mu\nu}=e^{~a}_{\mu}\cdot e^{~a}_{\nu}\,\eta_{ab}$, where $\eta_{ab}$ is the constant symmetric metric of flat Minkowski spacetime.  

With the conditions that the metric is a covariant tensor of rank two, that its inverse $G^{\mu\nu *}$ is also a tensor of rank two such that $G_{\mu\nu}*G^{\nu\rho *}=\delta_{\mu}^{\rho}$, and that the covariant derivative of $G_{\mu\nu}$ is zero, the other elements like Christoffel symbols, covariant derivative, Riemann curvature tensor, Ricci tensor and curvature scalar are constructed in terms of the metric tensor and the inverse metric tensor \cite{Aschieri:2005yw}. Our idea here is not to review the entire formulation of twist-deformed diffeomorphism algebra, but to point out that if $\Theta^{\mu\nu}$ is taken to be a constant and $\{x^\mu\}$ is, for example, simply the spherical polar coordinate set $(r,\theta,\phi,t)$ or the parabolic coordinate set $(u,v,\phi,t)$, then the deformation products with such twists also satisfy the associative property, and all of the above geometrical elements can be constructed in the spherical polar or parabolic coordinate systems as well along the lines described in \cite{Aschieri:2005yw}. 
\section{\label{sec:cartcyl}Cartesian and Cylindrical Coordinates in NC Case}
The convention we adopt is $x^{\mu}=(x^1,x^2,x^3,x^4)$ in which the spatial indices take the values $1,2$ and $3$. The  signature of $\eta_{ab}$ is taken to be $(+,+,+,-)$. 

When the tetrads are constants, it is clear from \eqref{eq-ncmetric} that the metric tensor $G_{\mu\nu}=g_{\mu\nu}$ to all orders in $\Theta$. Other geometrical elements are also not different from their commutative counterparts. The Cartesian system of coordinates falls under this category. In the case of Cartesian coordinates, the star product of functions is cyclic under the integration over the entire spacetime.

If the tetrads depend on only one coordinate, then also  $e^{~a}_{\mu}*e^{~b}_{\nu}=e^{~a}_{\mu} \cdot e^{~b}_{\nu}$, and all the geometrical elements are the same as in the commutative theory, because any nontrivial $\Theta$-corrections require the basic element $e^{~a}_{\mu}$ to be dependent on at least two distinct coordinates such that the corresponding $\Theta^{\alpha\beta}\neq 0$. An example for this type is the cylindrical system of coordinates $(\rho,\phi,z,t)$. The metric in this case is $g_{\mu\nu}=\mathrm{diag}(1,\rho^2,1,-c^2)$, and the  Drinfel'd twist  $e^{\frac{i\Theta^{\mu\nu}}{2}\frac{\partial}{\partial x^\mu} \otimes \frac{\partial}{\partial x^\nu}}$, where $\{x^\mu\}$ is the set $(\rho,\phi,z,t)$ and $\Theta^{\mu\nu}$ is a constant matrix, will not alter any geometrical element. However, the star product of functions under integration over cylindrical coordinates is not cyclic since the volume element depends on the polar coordinate $\rho$. But this non-cyclic nature can be rectified if we take $\Theta^{\rho\phi}=\Theta^{\rho z}=\Theta^{\rho t}=0$ keeping $\Theta^{\phi z}\neq 0$, $\Theta^{\phi t}\neq 0$ and $\Theta^{zt}\neq 0$, where $\rho$ is the polar coordinate. 

In the above two cases, only the interactions with other fields can reveal the underlying NC structure of spacetime. 
\section{\label{sec:tetrads}Tetrads with more than One Coordinate Dependency}
If the vierbein depends only on $x^1$ and $x^2$  and if these two coordinates commute, then in the pure gravity sector, there are no NC corrections even if the 
spacetime has other noncommuting coordinate pairs. Because all the basic 
geometric elements are constructed out of the *-product of tetrads as in 
Eq.\eqref{eq-ncmetric} and
\begin{align}
e^{~a}_{\mu}(x^1,x^2)*e^{~b}_{\nu}(x^1,x^2)= e^{~a}_{\mu} e^{~b}_{\nu} ~~\mathrm{if}~~ \Theta^{12}=0.
\end{align}
So the metric tensor will have no $\Theta$-dependency. The inverse metric, which has an expansion in $\Theta$ that depends only on the derivatives of $e^{~a}_{\mu}$ in its expressions for $\Theta$-corrections \cite{Aschieri:2005yw}, will also have no $\Theta$-corrections. The reason is both derivatives with respect to $x^{\alpha}$ and $x^{\beta}$ appear as multiplicative factors in each of the terms involving $\Theta^{\alpha\beta}$, and it is either that for nontrivial derivative $\Theta^{\alpha\beta}$ is zero or that for nontrivial $\Theta^{\alpha\beta}$ the derivative is zero, making each term to vanish. So the metric and the inverse metric will be the same as their commutative counterparts. By the same token, the Christoffel symbols and Riemann tensor will also be the same as their commutative counterparts. Only the interaction with other gauge or matter fields, which in general depend on all coordinates, can reveal such underlying noncommuting structure of coordinates if $\Theta^{\alpha\beta}\neq 0$ for other pairs of distinct values of $\alpha$ and $\beta$. 

\emph{ But if the vierbein $e^{~a}_{\mu}$ is a function of, for example,   $x^1$ and $x^2$, and if these two coordinates are promoted as noncommuting operators, then nontrivial NC curvature corrections do show up.}
\section{Remarks on NC Scalar Invariants}\label{sec:scalarinv}
The nontrivial NC corrections we will work out have singularities in certain coordinate limits. In the commutative case, a singularity is not a coordinate singularity when the higher-order scalar invariants remain singular at certain points (See, for example, \cite{Carroll:2004st}). In the same way, to corroborate that the singularities in the NC cases are not coordinate singularities, we focus our attention on an NC version of the Kretschmann scalar. Since the NC corrections to Riemann tensor elements are in general complex-valued, we construct the following real-valued NC Kretschmann scalar and write it in the  way that the real nature is manifest:
\begin{align}
K= R_{\mu\nu\rho\sigma}*G^{\sigma\alpha *}*G^{\rho\beta *}*G^{\nu\gamma *}*G^{\mu\delta *}* \bar{R} _{\alpha\beta\gamma\delta}, \label{2nd-order-kretschmann}
\end{align}
where $\bar{R}_{\alpha\beta\gamma\delta}$ is the complex conjugate of $R_{\alpha\beta\gamma\delta}$. In the commutative limit it reduces to $K^{(0)}=R^{(0)}_{\mu\nu\rho\sigma} R^{(0)\sigma\rho\nu\mu}=R^{(0)}_{\mu\nu\rho\sigma}R^{(0)\mu\nu\rho\sigma}$.

In cases in which $R^{(0)}_{\mu\nu\rho\sigma}$=0, the first order correction $K^{(1)}$ is also zero, and there is only one nontrivial second-order term: 
\begin{align}
K^{(2)}= R^{(1)}_{\mu\nu\rho\sigma}\,g^{\sigma\alpha}\,g^{\rho\beta}\,g^{\nu\gamma }\, g^{\mu\delta} \,\bar{R}^{(1)}_{\alpha\beta\gamma\delta}.
\end{align}
In addition to the above NC correction, we will calculate other geometrical elements in the following cases. 
\section{NC Minkowski Spacetime in Spherical Polar Coordinates}\label{sec:ncmspheri}
The reason why this coordinate system $(r,\theta,\phi,t)$ is chosen for the curvature analysis is that the tetrad depends on more than one coordinate i.e., $r$ and $\theta$. In the commutative case, the tetrad, the metric tensor and the inverse of the metric tensor in this case are given by
\begin{align}
\begin{aligned}
e^{~a}_{\mu}&= \mathrm{diag}(1,\,r,\,,r\sin\theta,\,-c); \\ 
g_{\mu\nu}&=\mathrm{diag}(1,\,r^2,\,r^2\sin^2\theta,\,-c^2);
\\g^{\mu\nu}&=\! \mathrm{diag}(1,\,r^{-2},\,(r\sin\theta)^{-2},\,-c^{-2}). \label{eq-cmetric}
\end{aligned}
\end{align}
In the commutative case, the non-zero  
Christoffel symbols of the second kind are given by (see for example 
\cite{Muller:2009bw}):
\begin{subequations}\label{eq-zo-chris}
\begin{alignat}{5}
  \Gamma_{\phi\phi}^{(0)}{}^{r} &= -r\sin^2\theta, & ~\Gamma_{\theta\theta}^{(0)}{}^{r} &= -r, & ~~ 
  \Gamma_{r\theta}^{(0)}{}^{\theta} &= \frac{1}{r}, \label{eq-zo-chris1}\\
  \Gamma_{\phi\phi}^{(0)}{}^{\theta} &= -\sin\theta\cos\theta, & ~~
  \Gamma_{r\phi}^{(0)}{}^{\phi} &= \frac{1}{r}, & 
  \Gamma_{\theta\phi}^{(0)}{}^{\phi} &= \cot\theta, \label{eq-zo-chris2}
\end{alignat}
\end{subequations}
with the proviso that $\Gamma_{\mu\nu}^{(0)}{}^{\rho}=\Gamma_{\nu\mu}^{(0)}{}^{\rho}$. The Riemann tensor $R_{\mu\nu\rho}^{(0)}{}^{\sigma}$, the Ricci tensor $R_{\mu\nu}^{(0)}$ and the curvature scalar $R^{(0)}$ are all zero. 

As stated at the end of Section \ref{sec:ncintro}, when the coordinates are promoted as noncommuting coordinates through the Drinfel'd twist  
\begin{align}
\mathcal{F}= e^{\frac{i\Theta^{\mu\nu}}{2}\frac{\partial}{\partial x^\mu} \otimes \frac{\partial}{\partial x^\nu}},  \label{drinfeldspherical}
\end{align}
where $\{x^\mu\}$ is the set $(r,\theta,\phi,t)$ and $\Theta^{\mu\nu}$ is a constant matrix, the formalism of \cite{Aschieri:2005yw} is applicable in the construction of geometrical elements because the associative nature of star-product holds in this case. 

Since the tetrad depends only on $r$ and $\theta$, the only noncommutation relation that will lead to NC curvature effects is
\begin{align}
 [\hat{r},~\hat{\theta}]=i\Theta^{r\theta}. 
\end{align}
No other nontrivial $\Theta^{\mu\nu}$ will result in any change in the nature of curvature of the spacetime. Note that the physical dimension of $\Theta^{r\theta}$ is the same as length. If $\Theta^{r\theta}=0$, then the NC Minkowski spacetime is flat regardless of the presence of any other nontrivial $\Theta^{\mu\nu}$ in the theory.
\subsection{First-Order NC Corrections}
\subsubsection{Inverse Metric Tensor and Christoffel Symbols}
The metric tensor in the NC case, Eq.\eqref{eq-ncmetric}, is symmetric and real by its construction. Therefore, there is no first-order correction to it.

The first-order NC correction to the inverse of the metric tensor is given by \cite{Aschieri:2005yw}
\begin{align}
\begin{aligned}
G^{(1)\mu\nu *}=-\frac{i}{2}\Theta^{\alpha\beta}\partial_{\alpha}\left(g^{
\gamma\mu } \right)\partial_{\beta}\left(g_{\delta\gamma}\right)g^{\delta\nu}. \label{eq-foinmetric}
\end{aligned}
\end{align}
Substituting for the zeroth order metric and the inverse metric components from Eq.\eqref{eq-cmetric} into Eq.\eqref{eq-foinmetric} gives $G^{(1)\mu\nu *}$=0, i.e., there are no first-order corrections to the inverse metric. 

The Christoffel symbols of the second kind have the following expression for their first-order correction:
\begin{align}
 \Gamma_{\mu\nu}^{(1)\rho}=\frac{i}{2}\Theta^{\alpha\beta}\left(\partial_{\alpha}\Gamma_{\mu\nu}^{(0)\sigma}\right)g_{\sigma\tau}(\partial_{\beta}g^{\rho\tau}), \label{eq-fo-chris}
\end{align}
and these are symmetric under the exchange of $\mu$ and $\nu$. For the spacetime under consideration, the independent non-zero symbols are calculated to be
\begin{align}
\begin{aligned}
 \Gamma_{\phi\phi}^{(1)\theta}&=i\Theta^{r\theta} \frac{(1-2\cos^2\theta)}{r}, \qquad
 \Gamma_{r\phi}^{(1)\phi}&=i\Theta^{r\theta} \frac{\cos\theta}{r^2\sin\theta}, &\qquad
 \\\Gamma_{\theta\phi}^{(1)\phi}&=-i\Theta^{r\theta}\frac{1}{r\sin^2\theta}, \label{eq-fo-chris-spher}
 \end{aligned}
\end{align}
i.e., only the zeroth order symbols corresponding to Eq.\eqref{eq-zo-chris2} have first-order $\Theta$-corrections and they vanish at the asymptotic limit $r\rightarrow \infty$. 
%
\subsubsection{Curvature Tensors and Curvature Scalar}
The NC version of Riemann tensor gives rise to the following first-order expression 
\begin{align}
\begin{aligned}
R_{\mu\nu\rho}^{(1)}{}^{\sigma}
\!=\!& -\frac{i}{2}\Theta^{\kappa\lambda}\bigg( (\partial_{\kappa}R_{\mu\nu\rho}^{(0)}{}^{\tau})
(\partial_{\lambda}g_{\tau\gamma})g^{\gamma\sigma} \bigg) \\ & + \frac{i}{2}\Theta^{\kappa\lambda}\bigg( 
(\partial_{\kappa}\Gamma_{\nu\rho}^{(0)}{}^{\beta})\Big(
\Gamma_{\mu\beta}^{(0)}{}^{\tau}(\partial_{\lambda}g_{\tau\gamma})g^{\gamma\sigma} \\
& -\Gamma_{\mu\tau}^{(0)}{}^{\sigma}(\partial_{\lambda}g_{\beta\gamma})g^{\gamma\tau}
 +\partial_{\mu}\big( (\partial_{\lambda}g_{\beta\gamma})g^{\gamma\sigma}\big) \\
& +(\partial_{\lambda}\Gamma_{\mu\beta}^{(0)}{}^{\sigma})\Big)  -\left(\mu\leftrightarrow\nu \right)\bigg).
\label{eq-fo-riemann}
\end{aligned}
\end{align}

It is antisymmetric under the exchange of $\mu$ and $\nu$. There are as many as 14 non-zero components, and 7 of them are independent for the spacetime under consideration. They are the following:
\begin{align}
R_{r\theta\phi}^{(1)}{}^{\phi}&\!=\!\frac{-3\,i\,\Theta^{r\theta}}{r^2\sin^2\theta},
& R_{r\phi\theta}^{(1)}{}^{\phi}&\!=\!\frac{i\,\Theta^{r\theta}(\cos^2\theta-2)}{r^2\sin^2\theta}, \nonumber\\
R_{r\phi\phi}^{(1)}{}^{\theta}& \!=\!\frac{i\,\Theta^{r\theta}(2-3\cos^2\theta)}{r^2}, 
& R_{\theta\phi r}^{(1)}{}^{\phi}&\!=\!\frac{i\,\Theta^{r\theta}(1+\cos^2\theta)}{r^2(1-\cos^2\theta)}, \nonumber \\
R_{\theta\phi\theta}^{(1)}{}^{\phi} & \!=\! \frac{i\Theta^{r\theta}\cot\theta}{r},
& R_{\theta\phi\phi}^{(1)}{}^{r} & \!=\! -i\Theta^{r\theta}\sin^2\theta , \nonumber\\
R_{\theta\phi\phi}^{(1)}{}^{\theta} & \!=\! \frac{i\Theta^{r\theta}\sin\theta\cos\theta}{r} \label{eq-fo-riemann-spher}. 
\end{align}
The six independent components of first-order NC correction of the Riemannian curvature tensor are inversely proportional to powers of $r$ and/or inversely proportional to powers of $\sin\theta$. The component $R_{\theta\phi\phi}^{(1)}{}^{r}$ is directly proportional to powers of $\sin\theta$. So this component does not vanish at the asymptotic limit $r\rightarrow \infty$. 

The first-order correction to the Ricci tensor is simply given by the sum over two indices in the first-order Riemann tensor, i.e., $ R_{\mu\nu}^{(1)}= R_{\mu\sigma\nu}^{(1)}{}^{\sigma} $, and it has 4 non-zero components:
\begin{align}
R_{r\theta}^{(1)}& \!=\! \frac{i\, \Theta^{r\theta}(\cos^2\theta-2)}{r^2\,\sin^2\theta }, & 
R_{\theta r}^{(1)} &\!=\!\frac{i\, \Theta^{r\theta}(1+\cos^2\theta)}{r^2\,(1-\cos^2\theta)},\nonumber \\
R_{\phi\phi}^{(1)} &\!=\!-i\, \Theta^{r\theta}\,\frac{\sin\theta\cos\theta}{r}, &
R_{\theta\theta}^{(1)}& \!=\! i\, \Theta^{r\theta}\,\frac{\cot\theta}{r}. \label{eq-fo-ricci-spher} 
\end{align}
The components of the first-order NC corrections to the Ricci tensor are inversely proportional to the power(s) of $r$. These components vanish at the radial asymptotic limit. The first-order correction to the curvature scalar is, however, zero. This can be easily seen since in our case $R^{(1)}=G^{\mu\nu * (1)}R_{\nu\mu}^{(0)}+g^{\mu\nu}R_{\nu\mu}^{(1)}+\tfrac{i}{2}\Theta^{\alpha\beta}(\partial_{\alpha}g^{\mu\nu})(\partial_{\beta}R_{\nu\mu}^{(0)})= g^{\theta\theta}R_{\theta\theta}^{(1)}+g^{\phi\phi}R_{\phi\phi}^{(1)}=0$. 
As it is evident from the above mathematical expressions, all the components except one of the first-order NC corrections to Riemannian curvature tensor and all the NC components of Ricci tensor in spherical polar coordinate vanish at the asymptotic limit $r\rightarrow \infty$.
\subsection{Second Order NC Corrections}
\subsubsection{Metric, Inverse Metric and Christoffel Symbols}
Because of the nature of the definition Eq.\eqref{eq-ncmetric} of the metric tensor in the NC spacetime, it has the second-order correction: 
\begin{align}
G_{\mu\nu}^{(2)} = 
- \frac{1}{8}\theta^{\alpha_{1}\beta_{1}}\theta^{\alpha_{2}\beta_{2}}
(\partial_{\alpha_{1}}\partial_{\alpha_{2}}e_{\mu}^{\ a})(\partial_{\beta_{1}}\partial_{\beta_{2}}e_{\nu}^{\ b})
\eta_{ab} \, .\label{eq-so-metric}
\end{align}
For the case under consideration, the only non-zero component is
\begin{align}
 G_{\phi\phi}^{(2)}=\frac{1}{4}(\Theta^{r\theta})^2 \cos^2\theta. \label{eq-so-metric-spher}
\end{align}
The second-order correction to the metric tensor does not vanish at the asymptotic limit.
The second-order correction to the inverse metric tensor in the general case is given by:
%
\begin{align}
G^{(2)\mu\nu\star}\!= &
 \frac{1}{8}\theta^{\alpha_{1}\beta_{1}}\theta^{\alpha_{2}\beta_{2}}g^{\eta\nu}
\bigg(  (\partial_{\alpha_{1}}\partial_{\alpha_{2}}g^{\mu\gamma})(\partial_{\beta_{1}}\partial_{\beta_{2}}g_{\gamma\eta}) \nonumber \\
& ~ + g^{\mu\gamma}(\partial_{\alpha_{1}}\partial_{\alpha_{2}}e_{\gamma}{}^{a})
(\partial_{\beta_{1}}\partial_{\beta_{2}}e_{\eta}{}^{b})\eta_{ab} \label{eq-so-invmetric}\\
& ~ -2\partial_{\alpha_{1}}\left((\partial_{\alpha_{2}}g^{\mu\gamma})(\partial_{\beta_{2}}g_{\gamma\delta})
 g^{\delta\epsilon}\right)(\partial_{\beta_{1}}g_{\epsilon\eta})\bigg)  \nonumber,
\end{align}
%
and for the NC Minkowski spacetime in spherical polar coordinates, the only component that has nontrivial correction is 
\begin{align}
G^{(2)\phi\phi\star} =-\left(\Theta^{r\theta}\right)^2 \frac{(4+\cos^2\theta)}{4\,r^4\,\sin^4\theta}. \label{eq-so-invmetric-spher}
\end{align}
So second-order NC corrections to inverse metric tensor fall off as $r^{-4}$.
The Christoffel symbols of the second kind have a rather lengthy expression for its second-order correction \cite{Aschieri:2005yw} and we give here only the  independent non-zero components:
\begin{align}
\Gamma_{\phi\phi}^{(2)\theta}&=\frac{-13\,(\Theta^{r\theta})^2\sin2\theta}{8\,r^2} , ~~
 \Gamma_{r\phi}^{(2)\phi} =\frac{(\Theta^{r\theta})^2(2-5\cos^2\theta)}{4\,r^3\,\sin^2\theta},~~\nonumber\\
 \Gamma_{\theta\phi}^{(2)\phi} &=\frac{-3\,(\Theta^{r\theta})^2\cos\theta}{4\,r^2\,\sin^3\theta}. \label{eq-so-chris-spher}
\end{align}
So it turns out that only those Christoffel symbols that have first-order corrections acquire nontrivial second-order corrections as well. These corrections vanish at the asymptotic limit ${r\to\infty}$.
\subsubsection{Curvature Tensors}
The following general expression for the second-order correction to Riemann tensor involves only the Christoffel symbols and their derivatives,
\begin{align}
R_{\mu\nu\rho}^{(2)}{}^{\sigma}&\!\!=\partial_{\nu}\Gamma_{\mu\rho}^{(2)}{}^{\sigma}
+\Gamma_{\nu\rho}^{(2)}{}^{\gamma}\Gamma_{\mu\gamma}^{(0)}{}^{\sigma}
+\Gamma_{\nu\rho}^{(0)}{}^{\gamma}\Gamma_{\mu\gamma}^{(2)}{}^{\sigma} \nonumber\\&
 + \frac{i\theta^{\alpha\beta}}{2}\Big( (\partial_{\alpha}\Gamma_{\nu\rho}^{(1)}{}^{\gamma})(\partial_{\beta}\Gamma_{\mu\gamma}^{(0)}{}^{\sigma})
\!+\!(\partial_{\alpha}\Gamma_{\nu\rho}^{(0)}{}^{\gamma})(\partial_{\beta}\Gamma_{\mu\gamma}^{(1)}{}^{\sigma})
\Bigr) \nonumber \\&
-\frac{1}{8}\theta^{\alpha_{1}\beta_{1}}\theta^{\alpha_{2}\beta_{2}}
(\partial_{\alpha_{1}}\partial_{\alpha_{2}} \Gamma_{\nu\rho}^{(0)}{}^{\gamma})
(\partial_{\beta_{1}}\partial_{\beta_{2}}\Gamma_{\mu\gamma}^{(0)}{}^{\sigma})\nonumber \\
&-(\mu\leftrightarrow\nu) .\label{eq-so-riemann}
\end{align}
These are antisymmetric under the exchange of $\mu$ and $\nu$. On the whole, there are 18 non-zero components in the present case, and 9 of them are independent. They are as follows:
\begin{align}
\begin{aligned}
 R_{r\theta\phi}^{(2)}{}^{\phi}&\!=\!\frac{-(\Theta^{r\theta})^2\,\cos\theta}{r^3\,\sin^3\theta},
\\R_{r\phi r}^{(2)}{}^{\phi}& \!=\!\frac{9\,(\Theta^{r\theta})^2\,(2-5\cos^2\theta)}{4\,r^4\,\sin^2\theta},\nonumber\\
R_{r\phi\theta}^{(2)}{}^{\phi}& \!=\!\frac{-(\Theta^{r\theta})^2\,(5\cos^2\theta+16)\cos\theta}{4\,r^3\,\sin^3\theta},
\\R_{r\phi\phi}^{(2)}{}^{r}& =\frac{(\Theta^{r\theta})^2\,(7\,\cos^2\theta+2)}{4\,r^2}, \nonumber \\
R_{r\phi\phi}^{(2)}{}^{\theta}& =\frac{5\,(\Theta^{r\theta})^2\,(21\,\cos^2\theta-20)\cos\theta}{4\,r^3\,\sin\theta},
\\R_{\theta\phi r}^{(2)}{}^{\phi}& =\frac{-(\Theta^{r\theta})^2\,(5\,\cos^2\theta+12)\cos\theta}{4\,r^3\,\sin^3\theta}, \nonumber\\
R_{\theta\phi\theta}^{(2)}{}^{\phi}& =\frac{(\Theta^{r\theta})^2\,(5\,\cos^4\theta-21\cos^2\theta+1)}{4\,r^2\,\sin^4\theta},
\\R_{\theta\phi\phi}^{(2)}{}^{r}& =\frac{-(\Theta^{r\theta})^2\,(5\,\cos^2\theta+6)\cos\theta}{4\,r\,\sin\theta}, \nonumber\\
R_{\theta\phi\phi}^{(2)}{}^{\theta}& =\frac{(\Theta^{r\theta})^2 (55\cos^2\theta-43\cos^4\theta-15)}{4\,r^2\,\sin^2\theta}. 
\end{aligned}
\end{align}
The sum over the repeated index in $R_{\mu\sigma\nu}^{(2)}{}^{\sigma}$ leads to nontrivial second-order corrections to the following five Ricci tensor components:
\begin{align}
R_{rr}^{(2)} &\!=\!\frac{9(\Theta^{r\theta})^2\,(2-5\cos^2\theta)}{4\,r^4\,\sin^2\theta} , 
\nonumber\\R_{r\theta}^{(2)}&\!=\!\frac{-(\Theta^{r\theta})^2(5\cos^2\theta+16)\cos\theta}{4\,r^3\,\sin^3\theta},
\nonumber\\R_{\theta r}^{(2)} &\!=\!\frac{-(\Theta^{r\theta})^2\,(12+5\cos^2\theta)}{4\,r^3\,\sin^3\theta},
\label{eq-so-ricci-spher} \\R_{\theta\theta}^{(2)} &\!=\!\frac{(\Theta^{r\theta})^2\,(5\cos^4\theta-21\cos^2\theta+1)}{4\,r^2\,\sin^4\theta},
\nonumber\\R_{\phi\phi}^{(2)} &\!=\!\frac{\Theta^{r\theta})^2}{4\,r^2}\!\!\left[\frac{(43\cos^4\theta-55\cos^2\theta+15}{\sin^2\theta} -7\cos^2\theta-2\right]\! . \nonumber
\end{align}
\subsubsection{Scalar Invariants}
The curvature scalar in the second-order is given as 
\begin{align}
\begin{aligned}
R^{(2)} =~&
  G^{(2)\mu\nu\star}R_{\nu\mu}^{(0)} + g^{\mu\nu}R_{\nu\mu}^{(2)} + G^{(1)\mu\nu\star}R_{\nu\mu}^{(1)}\\ 
  &\!\!+\frac{i}{2}\theta^{\alpha\beta}(\partial_{\alpha}g^{\mu\nu})(\partial_{\beta}R_{\mu\nu}^{(1)})\\
 &-\frac{1}{8}\theta^{\alpha_{1}\beta_{1}}\theta^{\alpha_{2}\beta_{2}}
 (\partial_{\alpha_{1}}\partial_{\alpha_{2}}g^{\mu\nu})(\partial_{\beta_{1}}
 \partial_{\beta_{2}}R_{\mu\nu}^{(0)}). \label{eq-so-curvscalar}
\end{aligned}
\end{align}
Unlike the first-order, this second-order picks up nontrivial correction, and the result is
\begin{align}
 R^{(2)}=\frac{4\,(\Theta^{r\theta})^2(7\cos 4\theta-11\cos 2\theta-2)}{r^4\,(\cos 4\theta-4\,\cos 2\theta+3)}. \label{eq-so-curvscalar-spher}
\end{align}
The second order NC correction to Kretschmann scalar Eq.(\ref{2nd-order-kretschmann}) is calculated using the first order NC correction to Riemannian curvature tensor Eq.(\ref{eq-fo-riemann-spher}) and it is given by
\begin{align}
K^{(2)}=\frac{-16(\Theta^{r\theta})^2}{r^{10}\sin^2\theta} \label{eq-so-kretschmannscalar}
\end{align}
\subsection{Discussions on the above Corrections}
As we can see from the above expressions, 
there is only one component of noncommutative correction of Riemannian curvature tensor $R^{(1)}_{\theta\phi\phi}{}^r$ which does not vanish at the limit ${r\to\infty}$. All other components of first-order and second-order noncommutative corrections to Christoffel symbols, Riemannian curvature tensor, Ricci tensor and Ricci scalar vanish at the radial asymptotic limit. 

Second-order noncommutative correction to Kretschmann scalar has singularities at $r=0$, $\theta=0$ and $\theta=\pi$. These singularities are interpreted in the following way. Although the coordinates $(r,\theta,\phi,t)$ represent points in spacetime, they should be thought of as the experimentally measured values relating them to the size of the probe at time $t$. If we are using probes to make length measurements and gradually making the probes smaller and smaller to investigate the physics at the Planck length scale such that its Compton wavelength becomes equal to the Planck's length, then the radius of the probe becomes smaller than the Schwarszschild's radius for the mass and energy range of the probe used for the measurement. So, as $r\rightarrow 0$, the probe used for the measurement turns into a black hole. Also, as we gradually reduce the value of $\theta=S/r$, we are basically reducing the arc-length $S$, i.e., the distance between two points along the section of a curve while keeping the radius $r$ unchanged. At the singularity, $\theta\rightarrow 0$, the arc-length $S\rightarrow 0$, but the radius remains constant. So we have a singularity stretching from $r=0$ to $r=\infty$. Since $\sin^2(\pi\pm \theta)=\sin^2\theta$ in the Krethschmann scalar Eq.(\ref{eq-so-kretschmannscalar}), the arc-length $S\rightarrow 0$ in this case also resulting in a singlarity. This case is intriguing because it represents a periodicity behavior of the singularity corresponding to the size of the probe and it is peculiar to this coordinate system. So in NC Minkowski spacetime parameterized by spherical polar coordinates, due to the measurement process, the singular point $r=0$ mimics a black hole, and a singular region stretches from $r=0$ to $r=\infty$ at $\theta=0$ and $\theta=\pi$. Such singularities do not arise in the cases of Cartesian and cylindrical coordinate systems. 

Note that in the case of spherical coordinates, the volume element involves $r^2\sin\theta$, and so the spacetime integration over the star product of functions obtained through the twist Eq.(\ref{drinfeldspherical}) is not cyclic. But if all the distinct $\Theta^{\mu\nu}$'s are made zero except for $\Theta^{t\phi}$, then the cyclic property is obeyed. Since $\Theta^{r\theta}$ is made zero in this process, there won't be any NC curvature corrections, and all the singularities can be avoided. 
\section{NC Minkowski Spacetime in Parabolic Coordinates}\label{sec:ncmpara}
For the definition and discussion of this coordinate system in the 3-D commutative case, see for example \cite{moon1971field}. If we denote the Minkowski spacetime coordinates as $ x^{\mu}=(x^1,x^2,x^3,x^4)=(u,v,\phi,t),$ then the tetrad can be compactly specified by
\begin{align}
 e^{~a}_{\mu}=\mathrm{diag}\left(\sqrt{u^2+v^2},~ \sqrt{u^2+v^2},~ uv,~ -c\right), \label{eq-tetrad-para}
\end{align}
and the corresponding metric tensor and the inverse metric tensor are given by
\begin{align}
\begin{aligned}
\!\!g_{\mu\nu}=\mathrm{diag}\left(u^2+v^2\!,\, u^2+v^2\!,\, u^2v^2\!,\, -c^2\right)\!, \\  g^{\mu\nu}\!=\mathrm{diag}\left(\frac{1}{u^2+v^2},\frac{1}{u^2+v^2},\frac{1}{u^2v^2},\frac{-1}{c^2}\right). \label{eq-cmetric-para}
\end{aligned}
\end{align}
We choose this coordinate system because the corresponding tetrad is dependent on two coordinates. In addition to the symmetric nature of the metric, we also have $g_{uu}=g_{vv}$. This results in 6 distinct non-zero Christoffel symbols in the zeroth order case, and they are worked out to be
\begin{align}
\begin{aligned}
\Gamma_{uu}^{(0)}{}^{u}&=-\Gamma_{vv}^{(0)}{}^{u}=\Gamma_{uv}^{(0)}{}^{v}=\frac{u}{u^2+v^2},\\
\Gamma_{uv}^{(0)}{}^{u}&=-\Gamma_{uu}^{(0)}{}^{v}=\Gamma_{vv}^{(0)}{}^{v}=\frac{v}{u^2+v^2}, \\
\Gamma_{\phi\phi}^{(0)}{}^{u}&=\frac{-uv^2}{u^2+v^2},\\
\Gamma_{\phi\phi}^{(0)}{}^{v}&=\frac{-u^2v}{u^2+v^2},\\
\Gamma_{u\phi}^{(0)}{}^{\phi}&=\frac{1}{u},\\
\Gamma_{v\phi}^{(0)}{}^{\phi}&=\frac{1}{v}. \label{eq-zo-chris-para}
\end{aligned}
\end{align}
But in this commutative case, the Riemann tensor $R_{\mu\nu\rho}^{(0)}{}^{u}$, the Ricci tensor $R_{\mu\nu}^{(0)}$ and the curvature scalar $R^{(0)}$ are all zero. 

Let's promote the two coordinates $u$ and $v$ as noncommuting operators with the commutation relation
\begin{align}
 [\hat{u},~\hat{v}]=i\Theta^{uv},  \label{eq-nc-commu-para}
\end{align}
via the twist  $e^{ \frac{i}{2}\Theta^{\mu\nu}\frac{\partial}{\partial x^{\mu}}\otimes\frac{\partial}{\partial x^{\nu}}}$, where  $(x^1,x^2,x^3,x^4)=(u,v,\phi,t),$ and $\Theta^{\mu\nu}$ is a constant matrix. The twist, in addition to Eq.(\ref{eq-nc-commu-para}), leads to other nontrivial commutation results in the theory, but the other commutation relations are not of our concern as they won't be changing the curvature properties of the geometry. 
\subsection{First Order NC Corrections}
\subsubsection{Inverse Metric Tensor and Christoffel Symbols}
Substitution of the zeroth order metric components Eq.\eqref{eq-cmetric-para} into Eq.\eqref{eq-foinmetric} results in $G^{(1)\mu\nu *}=0$, i.e., there are no first-order corrections to the inverse metric tensor. 

The first-order non-zero Christoffel symbols Eq.\eqref{eq-fo-chris} are worked out to be 
\begin{align}
\begin{aligned}
\Gamma_{uu}^{(1)}{}^{u}&=\frac{-i\Theta^{uv} \,v}{(u^2+v^2)^2}= \Gamma_{uv}^{(1)}{}^{v} = 
\Gamma_{vv}^{(1)}{}^{u}, \\
\Gamma_{uu}^{(1)}{}^{v}&=\frac{-i\Theta^{uv} \,u}{(u^2+v^2)^2} =\Gamma_{vv}^{(1)}{}^{v} =-\Gamma_{uv}^{(1)}{}^{u}, \\
\Gamma_{u\phi}^{(1)}{}^{\phi}&=\frac{i\Theta^{uv}}{u^2v}, \quad \qquad
\Gamma_{v\phi}^{(1)}{}^{\phi} =\frac{-i\Theta^{uv}}{uv^2}, \\
\Gamma_{\phi\phi}^{(1)}{}^{u}&=\frac{i\Theta^{uv}\,v\,(v^2-2u^2)}{(u^2+v^2)^2}, \\
\Gamma_{\phi\phi}^{(1)}{}^{u}&=\frac{i\Theta^{uv}\,v\,(v^2-2u^2)}{(u^2+v^2)^2}. \label{eq-fo-chris-para}
\end{aligned}
\end{align}
\subsubsection{Curvature Tensors and Curvature Scalar}
There are seven distinct non-zero components of the Riemann tensor in the first-order:
\begin{align}
\begin{aligned}
R_{uvu}^{(1)}{}^{u}&=R_{uvv}^{(1)}{}^{v}=\frac{2i\Theta^{uv}}{(u^2+v^2)^2},\\
R_{uv\phi}^{(1)}{}^{\phi}&=\frac{-3i\Theta^{uv}}{u^2v^2},\\
R_{u\phi v}^{(1)}{}^{\phi}&=\frac{-i\Theta^{uv}(u^2+2v^2)}{(u^2+v^2)u^2v^2},\\
R_{u\phi\phi}^{(1)}{}^{u}&=R_{v\phi\phi}^{(1)}{}^{v}=\frac{-4i\Theta^{uv}(u^2-v^2)uv}{(u^2+v^2)^3},\\
R_{u\phi\phi}^{(1)}{}^{v}&=\frac{-i\Theta^{uv}(u^2-7v^2)u^2}{(u^2+v^2)^3},\\ 
R_{v\phi u}^{(1)}{}^{\phi}&=\frac{i\Theta^{uv}(2u^2+v^2)}{(u^2+v^2)u^2v^2},\\
R_{v\phi\phi}^{(1)}{}^{u}&=\frac{-7i\Theta^{uv}\,v^2\,(u^2-v^2/7)}{(u^2+v^2)^3}. \label{eq-fo-riemann-para}
\end{aligned}
\end{align}
There are only three non-zero first-order components for the Ricci tensor:
\begin{align}
\begin{aligned}
 R_{uv}^{(1)}&=\frac{-i\Theta^{uv}(u^4+u^2v^2+2v^4)}{(u^2+v^2)^2 u^2v^2} , 
 \\R_{vu}^{(1)}&=\frac{i\Theta^{uv}(2u^4+u^2v^2+v^4)}{(u^2+v^2)^2 u^2v^2} , \\
 R_{\phi\phi}^{(1)}&=\frac{8i\Theta^{uv}(u^2-v^2)uv}{(u^2+v^2)^3}. \label{eq-fo-ricci-para}
\end{aligned}
\end{align}
Unlike the case of spherical polar coordinates, the parabolic system leads to nontrivial first-order correction to the curvature scalar:
\begin{align}
 R^{(1)}=\frac{8i\Theta^{uv}(u^2-v^2)}{(u^2+v^2)^3}. \label{eq-fo-curvscal-para}
\end{align}
\subsection{Second Order NC Corrections}
\subsubsection{Metric, Inverse Metric and Christoffel Symbols}
The only component of the metric tensor that picks up the second-order NC correction is
\begin{align}
 G_{\phi\phi}^{(2)}= \frac{(\Theta^{uv})^2}{4}. \label{eq-so-metric-para}
\end{align}
But there are three non-zero second-order components of the inverse metric tensor and all of them are diagonal:
\begin{align}
\begin{aligned}
G^{(2)uu*}&= \frac{(\Theta^{uv})^2}{(u^2+v^2)^3}, \qquad
G^{(2)vv*}&= \frac{(\Theta^{uv})^2}{(u^2+v^2)^3}, \qquad
\\G^{(2)\phi\phi*}&= \frac{-5(\Theta^{uv})^2}{u^4v^4}.
\end{aligned}
\end{align}
The Christoffel symbols pick up the following six distinct non-zero second-order corrections: 
\begin{align}
\begin{aligned}
 \Gamma_{uu}^{(2)}{}^{u} &=\frac{(\Theta^{uv})^2\,u}{(u^2+v^2)^3} =\Gamma_{uv}^{(2)}{}^{v}=-\,\Gamma_{vv}^{(2)}{}^{u},\\
 -\Gamma_{uu}^{(2)}{}^{v}&=\frac{(\Theta^{uv})^2\,v}{(u^2+v^2)^3}=\Gamma_{uv}^{(2)}{}^{u}=\Gamma_{vv}^{(2)}{}^{v},\\
  \Gamma_{u\phi}^{(2)}{}^{\phi}&= \frac{-3(\Theta^{uv})^2}{4u^3v^2}, \\
  \Gamma_{v\phi}^{(2)}{}^{\phi}& = \frac{-3(\Theta^{uv})^2}{4u^2v^3},\\
  \Gamma_{\phi\phi}^{(2)}{}^{u}&= \frac{(\Theta^{uv})^2\,u\,(3u^2-11v^2)}{2(u^2+v^2)^3},\\ 
  \Gamma_{\phi\phi}^{(2)}{}^{v}&= \frac{(\Theta^{uv})^2\,v\,(11u^2-3v^2)}{2(u^2+v^2)^3}.\\ 
\end{aligned}
\end{align}
\subsubsection{Curvature Tensors}
In NC Minkowski spacetime in parabolic coordinates, there are 11 Riemann tensor components that receive second corrections, apart from the antisymmetric ones under the exchange of the first two indices:
\begin{align}
R_{uvu}^{(2)}{}^{v}&=-\,R_{uvv}^{(2)}{}^{u}=\frac{12(\Theta^{uv})^2}{(u^2+v^2)^3},\nonumber\\
R_{u\phi v}^{(2)}{}^{\phi}&=\frac{-\,(\Theta^{uv})^2(29u^4+42u^2v^2+21v^4)}{4(u^2+v^2)^2 u^3v^3},\nonumber\\
R_{uv\phi}^{(2)}{}^{\phi}&=\frac{2\,(\Theta^{uv})^2(-u^2+v^2)}{(u^2+v^2)u^3v^3},\nonumber \\
R_{u\phi \phi}^{(2)}{}^{u}&=\frac{(\Theta^{uv})^2(101u^6-393u^4v^2+39u^2v^4-3v^6)}{4(u^2+v^2)^4 u^2},\nonumber\\
R_{u\phi u}^{(2)}{}^{\phi}&=\frac{-\,3(\Theta^{uv})^2(13u^2+5v^2)}{4(u^2+v^2)u^4v^2}, \nonumber \\
R_{u\phi \phi}^{(2)}{}^{v}&=\frac{(\Theta^{uv})^2(9u^6-341u^4v^2+175u^2v^4-11v^6)}{4(u^2+v^2)^4 uv},\nonumber\\
R_{v\phi v}^{(2)}{}^{\phi}&=\frac{-\,3(\Theta^{uv})^2(5u^2+13v^2)}{4(u^2+v^2)u^2v^4}, \nonumber \\
R_{v\phi u}^{(2)}{}^{\phi}&=\frac{-\,(\Theta^{uv})^2(21u^4+42u^2v^2+29v^4)}{4(u^2+v^2)^2u^3v^3}, \nonumber \\
%
R_{v\phi\phi}^{(2)}{}^{u}&= \frac{-\,(\Theta^{uv})^2(11u^6-175u^4v^2+341u^2v^4-9v^6)}{4(u^2+v^2)^4uv}~, \nonumber  \\
\end{align}
and
\begin{align}
R_{v\phi\phi}^{(2)}{}^{v}&= \frac{-\,(\Theta^{uv})^2(3u^6 -39u^4v^2+393u^2v^4-101v^6)}{4(u^2+v^2)^4 v^2}~. 
 \label{eq-so-riemann-para}
\end{align}
The sum over the repeated index in $R_{\mu\sigma\nu}^{(2)}{}^{\sigma}$ results in the second-order Ricci tensor, and there are five nontrivial components:
\begin{align}
\begin{aligned}
R_{uu}^{(2)}& =\frac{-\,3(\Theta^{uv})^2(13u^6 +15u^4v^2+23u^2v^4+5v^6)}{4(u^2+v^2)^3 u^4v^2}, \\
R_{uv}^{(2)}&= \frac{-\,(\Theta^{uv})^2(29u^4 +42u^2v^2+21v^4)}{4(u^2+v^2)^2 u^3v^3}, \\
R_{vu}^{(2)}&= \frac{-\,(\Theta^{uv})^2(21u^4 +42u^2v^2+29v^4)}{4(u^2+v^2)^2 u^3v^3}, \\
R_{vv}^{(2)}& =\frac{-\,3(\Theta^{uv})^2(5u^6 +23u^4v^2+15u^2v^4+13v^6)}{4(u^2+v^2)^3 u^2v^4}, \\
R_{\phi\phi}^{(2)}& =\frac{(3u^8-140u^6v^2+786u^4v^4-140u^2v^6+3v^8)}{4( \Theta^{uv})^{-2} (u^2+v^2)^4 u^2v^2}.
\label{eq-so-ricci-para}
\end{aligned}
\end{align}
\subsubsection{Curvature Scalars}
The correction to the scalar curvature is worked out to be
\begin{align}
 R^{(2)}=\frac{(\Theta^{uv})^2(3u^8-2u^6v^2-54u^4v^4-2u^2v^6+3v^8)}{(u^2+v^2)^4 u^4v^4}.
\end{align}
%
%
Second order component of noncommutative correction to Kretschmann scalar is
\begin{align}
    K^{(2)}=\frac{-8(\Theta^{uv})^2(u^4+34u^2v^2+v^4)}{u^2v^2(u^2+v^2)^{10}} \label{eq-so-kretschmann-para}
\end{align}
\subsection{Discussions}
From the above expressions, we can see that the first-order and second-order  NC corrections to metric tensor, inverse metric tensor, Christoffel symbol, curvature tensor and curvature scalar vanish at the asymptotic limits ${u\to\infty}$ and/or ${v\to\infty}$. So NC Minkowski spacetime parameterized by parabolic coordinate is asymptotically flat in these two limits. The second-order NC correction to Kretschmann Scalar has singularities at the points $u=0$ and $v=0$. 
The transformation 
\begin{align}
(u,v,\phi,t) \longrightarrow  (uv\cos\phi,\,uv\sin\phi, \,\tfrac{1}{2}(u^2-v^2),\,t) \label{para2cart}
\end{align}
from the parabolic system to the Cartesian system, implies that $z=(u^2-v^2)/2$, the measurement of $u$ is about the measurement of $z$ in the positive $z$-direction when the value of $v$ is constant, and measuring $z$ in the negative $z$-direction gives the measurement of $v$ when $u$ is held constant \cite{Margenau1943}. So the singularities at $u=0$ and $v=0$ can happen due to the size of the probe getting smaller and smaller along the $z$-direction. The smaller size of the probe in the other dimensions leading to singularities can similarly be explained using Eq.(\ref{para2cart}). In the case of parabolic coordinates, the volume element is proportional to $(u^2+v^2)u^2v^2$, and so the integration of the star product of functions over the spacetime is not cyclic. But if all the components of NC parameter $\Theta^{\mu\nu}$ except $\Theta^{t\phi}$ are zero, then the cyclic property is obeyed. As $\Theta^{uv}$ is zero in such a case,  all nontrivial corrections to curvatures in NC spacetime also vanish, ensuring a theory free from singularities due to probes. 

\section{Concluding Remarks}\label{sec:conclu}

In order to illustrate the consequent effect of noncommutative (NC) gravity theory, we applied to flat Minkowski spacetime the Drinfel'd twist $e^{\frac{i\Theta^{\mu\nu}}{2}\frac{\partial}{\partial x^\mu} \otimes \frac{\partial}{\partial x^\nu}}$, where $\{x^\mu\}$ is either the set $(r,\theta,\phi,t)$ of spherical polar coordinates or the parabolic coordinate set $(u,v,\phi,t)$, and $\Theta^{\mu\nu}$ is a constant real antisymmetric matrix in either of the case. The noncommutation among the coordinates  $\left[\hat{x}^{\mu},~\hat{x}^{\nu}\right]=i\Theta^{\mu\nu}$ emerges as a result of the twist. The effect of noncommutation among the spacetime coordinates manifests as nontrivial NC corrections to the Riemannian curvature tensor, Ricci tensor and Ricci curvature scalar respectively in NC Minkowski spacetime. It is worth remarking that the theories on these coordinate systems are not twisted-diffeomorphic to each other or to the theories involving Cartesian or cylindrical systems of coordinates. 
 
The components of the metric tensor in the NC spacetime depend on star products of tetrad components. When tetrads are constants, metric tensor components in NC spacetime are indistinguishable from those in commutative spacetime. The values of all other geometric elements also do not differ from their commutative counterparts in these cases. The Cartesian coordinate system has constant tetrads, and the star product of functions in this spacetime is cyclic. In cases where the tetrad depends only on one coordinate, curvature tensor and other geometric elements do not have any noncommutative corrections. Any nontrivial NC correction to the curvature tensor and curvature scalar requires the tetrad components to depend on at least two noncommuting coordinates. Tetrads in cylindrical coordinate systems depend on one coordinate, and so explicit NC corrections due to only noncommutative gravity/geometry do not show up in this case. Integration of star products over cylindrical coordinates is not cyclic as the volume element depends on the polar coordinate $\rho$. To ensure the cyclicity of the star product under integration, the values of the deformation parameters can be taken as $\Theta^{\rho\phi}=\Theta^{\rho z}=\Theta^{\rho t}=0$ while $\Theta^{\phi z}\neq 0$, $\Theta^{\phi t}\neq 0$ and $\Theta^{z t}\neq 0$. Note that the noncommutativity between a rectilinear coordinate and time has issues with unitarity and and UV-IR divergences \cite{Gomis:2000zz,Alvarez-Gaume:2001dfr,Chu:2002fe,Bahns:2002vm}. But the study of the consequences of noncommutativity between a curvilinear coordinate like $\phi$ and time has not yet appeared in the literature. 
  
In the cases where tetrad components depend on two commuting coordinates, there are no NC corrections to the geometric elements due to NC gravity, as both metric tensor and inverse metric tensor do not show any nontrivial corrections in these cases. Noncommutative curvature corrections only show up in cases where the tetrad component is a function of two noncommuting spacetime coordinates. For spherical polar coordinates, the tetrads depend on $r$ and $\theta$. Similarly, in parabolic coordinates, the vielbein depends on the coordinates $u$ and $v$, and the twist elevates these to noncommutative coordinates. In these two cases introducing noncommutativity among coordinates leads to the emergence of NC curvature corrections. The integral of star products over spherical polar coordinates and parabolic coordinates are not cyclic as the volume elements are proportional to $r^2sin\theta$ and $u^2v^2(u^2+v^2)$ respectively. If all components of NC deformation parameters are taken as zero except $\Theta^{t\phi}$, the star product remains cyclic. In such cases, since $\Theta^{r\theta}$ and $\Theta^{uv}$ are zero, NC curvature corrections disappear in both spherical polar coordinate and parabolic coordinate systems.
  
In general, if the metric tensor in the commutative case depends on more than one curvilinear coordinates, then the introduction of coordinate-noncommutativity among these coordinates can lead to different curvature properties. NC corrections to first-order Riemannian curvature tensor do not vanish at the radial asymptotic limit for NC Minkowski spacetime parameterized by spherical polar coordinates. NC corrections to all the geometric elements vanish in the asymptotic limits in NC Minkowski spacetime parameterized by parabolic coordinates. So, NC Minkowski spacetime in spherical polar coordinates is not asymptotically flat at the limit $r\rightarrow\infty$, whereas noncommutative Minkowski spacetime parameterized by parabolic coordinates is asymptotically flat in both of the limits $u\rightarrow\infty$ and $v\rightarrow\infty$. It shows that the asymptotic nature of NC Minkowski spacetime differs depending on the coordinate systems that parameterize it. 
  
Noncommutative corrections to curvature scalars in spherical polar coordinates show singular behavior as $r \rightarrow 0$, $\theta \rightarrow 0$, and $\theta \rightarrow \pi$. Second-order NC corrections to the Kretschmann scalar also have singularities at these points. Therefore these singularities are not coordinate singularities. These values of the coordinates are connected to experimentally measurable length scales and to the size of probes needed for the measurements. At $r\rightarrow 0$, the size of the probe required to measure the length scale becomes so small that its radius becomes smaller than the Schwarzschild radius, and the probe turns into a black hole. At $\theta \rightarrow 0$ and $\theta \rightarrow \pi$, the singularity is a region that stretches from $r=0$ to $r=\infty$. In this limit, for any given value of the radius $r$, the arc length i.e., the distance between two points along an arc 
 tends to zero.
 
The second-order noncommutative corrections to the curvature scalar and Kretschmann scalar show singular behavior at the points $u=0$ and $v=0$ in the parabolic coordinate system. Measurements of $u$ and $v$ are connected to the length measurement along positive and negative directions of the $z$-axis, respectively. So the singularities at these points show up as the size of the probes to measure the length scale gets smaller and smaller gradually. So the singularities in this coordinate system can also be connected to probes turning into black holes at high energy limits and smaller length scales.

\section*{Acknowledgments}
MR would like to thank the DST for the INSPIRE fellowship. Thanks are also due to Dr.RSK for his efforts to develop the computational lab in the department.

\bibliographystyle{}


\end{document}